\documentstyle[twocolumn,prl,aps]{revtex}

\newcommand{\be}{\begin{equation}}
\newcommand{\ee}{\end{equation}}
\newcommand{\lb}{\label}
\newcommand{\en}{\epsilon}
\newcommand{\bc}{{\bf c}}
\newcommand{\bk}{{\bf k}}
\newcommand{\bj}{{\bf j}}
\newcommand{\wbj}{{\bf j}}
\newcommand{\br}{{\bf r}}
\newcommand{\bx}{{\bf x}}
\newcommand{\by}{{\bf y}}
\newcommand{\bD}{{\bf D}}
\newcommand{\bE}{{\bf E}}

\newcommand{\bL}{{\bf L}}
\newcommand{\bZ}{{\bf Z}}
\newcommand{\wbJ}{{\widehat{{\bf J}}}}

\newcommand{\wS}{{\widehat{S}}}
\newcommand{\wrho}{{\widehat{\rho}}}
\newcommand{\bdot}{{\mbox{\boldmath $\cdot$}}}
\newcommand{\grad}{{\mbox{\boldmath $\nabla$}}}
\newcommand{\bsigma}{{\mbox{\boldmath $\sigma$}}}
\newcommand{\bddot}{{\bf :}}
\newcommand{\bmu}{{\mbox{\boldmath $\mu$}}}

\begin{document}

\relax

\draft

\title{Shape-Dependent Thermodynamics and Non-Local Hydrodynamics in a
Non-Gibbsian
       Steady-State of a Drift-Diffusion System}
\author{Francis J. Alexander\\{\em Center for Computational Science, Boston
University, Boston, MA 02215}
and\\
Gregory L. Eyink\\{\em Department of Mathematics, University of Arizona,
Tucson, AZ 85721}}
\date{\today}
\maketitle

\begin{abstract}
Shape-dependent thermodynamics and non-local hydrodynamics are argued to occur
in dissipative steady states
of driven diffusive systems. These predictions are confirmed by numerical
simulations. Unlike power-law
correlations, these phenomena cannot be explained by a hypothesis of
``criticality''. Instead, they require
the effective Hamiltonian of the system to contain very long-range potentials,
making the invariant probability
measures formally ``non-Gibbsian''.
\end{abstract}

\narrowtext

Power-law decay of correlations are generic in dissipative steady-states of
open, driven systems with conservation
laws \cite{GLMS,GLS}. In equilibrium systems such power laws can be due to two
different mechanisms: the interaction
potentials in the system Hamiltonian can themselves be long-range power-laws
(e.g. dipolar) or else the potentials can be
short-ranged but the system may be at a critical point \cite{DS}.  The latter
circumstance has prompted the view that
dissipative, nonequilibrium systems are attracted without any tuning of
parameters to a critical state, or exhibit
so-called ``self-organized  criticality'' (SOC) \cite{BTW}. We shall show here
that such an interpretation, taken literally,
is not true in an important class of such systems. As we shall explain, an {\em
effective Hamiltonian} may be introduced
to characterize the nonequilibrium statistics. We then exhibit phenomena that
can be explained {\em only} by the presence
there of effective power-law potentials of very long range. Such an explanation
for power-law correlations was earlier
proposed in {\em driven diffusive systems} \cite{ZS} and, more recently, in a
shaken {\em granular flow}
\cite{Will}. Because the underlying dynamics are local in these systems, this
mechanism may be justly
termed ``self-organized long-range interactions'' (SOLRI). We shall demonstrate
in the class of systems considered
two closely related phenomena: {\em shape-dependent thermodynamics} and {\em
non-local hydrodynamics}. These
phenomena could not occur if the effective Hamiltonian were short-ranged but
critical. In fact, the induced
potentials must have such an extreme long-range, many-body character that the
nonequilibrium measures are
formally {\em non-Gibbsian}. (For an excellent review of the relevant notions,
see \cite{EFS}.)
The indicators of the non-Gibbsian nature, shape-dependent thermodynamics and
non-local hydrodynamics,
have great importance in themselves. They are expected to occur in physical
drift-diffusion systems,
e.g. electrolytes and semiconductors.

Correlations of power-law type are predicted in such systems by linear
fluctuating hydrodynamics \cite{GLMS,GLS}.
In a DDS with one species of (unit-charged) particle at constant mean density
$\rho$, in contact with a heat bath
at temperature $T$ and in an applied electric field $\bE$, the equation in
Fourier space for density
fluctuations $\delta \wrho(\bk,t)$ is
\be \delta \dot{\wrho}(\bk,t) =
[i\bc_\bE\bdot\bk-\bk\bdot\bD_\bE\bdot\bk]\delta \wrho(\bk,t)  -i\bk\bdot\delta
\wbJ(\bk,t). \lb{lin-eq} \ee
Here $\bc_\bE(\rho)={{d\bj_\bE}\over{d\rho}}(\rho)$ is the drift velocity, a
density derivative
of the conduction current. The latter is given, at low density $\rho$ and small
field-strength $E$, by Ohm's law,
$\bj_\bE=\bsigma\bdot\bE=\rho\bmu\bdot\bE$, with $\bsigma$ the conductivity
tensor and $\bmu$ the mobility tensor.
$\bD_\bE(\rho)$ is the diffusion tensor in the conducting state, while $\delta
\wbJ(\bk,t)$ is the Fourier component
of the current noise with zero mean and covariance $\langle \delta
\wbJ(\bk,t)\delta \wbJ(\bk',t')\rangle=
2\bL_\bE^s(\rho)\delta^3(\bk+\bk') \delta(t-t')$. In the limit $E\rightarrow
0$, $\bL_\bE(\rho)$ coincides with
the Onsager matrix $\bL=\bsigma k_BT$. It is easy to solve Eq.(\ref{lin-eq})
for the static structure function
$\wS_\bE(\bk)\equiv \lim_{t\rightarrow\infty}\langle
\delta \wrho(\bk,t)\delta \wrho(-\bk,t)\rangle$:
\be \wS_\bE(\bk)= {{\bk\bdot\bL_\bE\bdot\bk}\over{\bk\bdot\bD_\bE\bdot\bk}}.
\lb{stat-struc} \ee
See \cite{GLMS}. Inverse Fourier transforming gives a long-range
density-density correlation $S_\bE(\br)\sim r^{-d}$
in dimension $d$, when $\bL_\bE$ and $\bD_\bE$ are not proportional. This will
generally be true for $E\neq 0$. Such
power-law correlations have been verified in numerical simulations of simple
{\em driven lattice gas} (DLG) models
\cite{KLS,ZWLV}. Correlations of this type are expected more generally in
locally driven/damped conservative systems
without detailed balance. For example, $r^{-d}$ correlations of similiar origin
have recently been predicted and verified
computationally in the homogeneous cooling state of rapid granular flows
\cite{OBNE}. Our results have implications
for all such systems.

We illustrate our points with the DLG models \cite{KLS,SZ}. These were
originally introduced as models of solid
electrolytes, or superionic conductors \cite{SK,DFP}. However, unlike physical
drift-diffusion systems of
charged particles, these models have only short-ranged dynamical interactions
and are thus perfect for
our theoretical objectives. We shall comment later how the results carry over
to realistic DDS with Coulombic
interactions. The particles of the DLG model live on a cubic lattice $\bZ^d$.
Assuming hard-core exclusion,
the occupancy $\eta_{\bx,t}\in\{0,1\}$ for each site $\bx\in\bZ^d$ and time
$t\geq 0$. The evolution of the configuration
is via a Kawasaki exchange dynamics, specified by the rate
$c_\bE(\bx,\by;\eta)$ for exchange of occupancy
of nearest neighbor sites $\bx,\by$ in the configuration $\eta$, i.e. for the
transition $\eta\rightarrow \eta^{\bx\by}$.
In  addition to assuming that rates are functions of occupancies at sites
within a finite range of $\{\bx,\by\}$,
the main assumption is {\em local} detailed balance:
\begin{eqnarray}
c_\bE(\bx,\by;\eta) & = & c_\bE(\bx,\by;\eta^{\bx\by})\exp\left[-\beta\left(
H(\eta^{\bx\by})-H(\eta) \right.\right.\cr
       \, &  &
\,\,\,\,\,\,\,\,\,\,\,\,\,\,\,\,\,\,\,\,\,\,\,\,\,\,\,\,\,\,\,\,\,\,\,\,\,
               \left.\left.
+\bE\bdot(\bx-\by)(\eta_\bx-\eta_\by)^{\,}\right)\right], \lb{ldb}
\end{eqnarray}
for some short-ranged lattice-gas Hamiltonian $H(\eta)$, e.g. an Ising model.
The condition (\ref{ldb}) encourages
particles to hop in the direction of the electric field $\bE$ and, in infinite
volume, sets up an irreversible
steady-state with a mean current. In fact, these models have space-ergodic,
homogeneous, and time-invariant measures
$\mu_{\rho,\bE,\beta}$ for each density $\rho\in[0,1]$, expected to be unique
at small $\beta$. The question arises
whether these measures in infinite volume are ``Gibbsian'' for an effective
Hamiltonian
\be H_{{\rm eff}}(\eta) = \sum_{A\subset\bZ^d} \Phi_A(\eta) \lb{Ham} \ee
with some set of many-body potentials $\Phi_A$ depending on spins $\eta_\bx$ at
sites $\bx\in A\subset\bZ^d$.
If it exists, this will generally not be the same as the short-ranged
Hamiltonian $H(\eta)$ used in defining the dynamics.
It turns out that almost {\em any} reasonable measure (with local densities) is
``Gibbsian'' if one permits
extremely long-ranged, many-body potentials and, indeed, there are a myriad of
physically inequivalent
such Hamiltonians! (E.g. see \cite{Israel}, Theorem V.2.2(a)). To guarantee
uniqueness and other standard
properties of usual Gibbs measures, the condition of {\em absolute summability}
is required:
\be \sum_{A\ni\bx}\|\Phi_A\|_\infty < \infty \lb{abs-sum} \ee
for all $\bx\in\bZ^d$, where $\|\Phi_A\|_\infty\equiv \sup_\eta
|\Phi_A(\eta)|.$ Following rather
common practice \cite{EFS}, we shall agree here to call only measures with the
latter property ``Gibbsian''.
It is a rigorous theorem of Asselah \cite{Assel} (or Appendix B of \cite{ELS})
that the following
alternative holds: either {\em all} space-ergodic, invariant measures of the
DLG are Gibbsian with
absolutely summable potential or else {\em none} of them are. It may also be
proved that in the domain
of analyticity, no absolutely summable power-law potential can produce a
correlation $\propto r^{-d}$
(Theorem 1 of \cite{DS}), such as is observed in the DLG. Thus, under the first
alternative, one is led to
conclude that the Gibbs measure must be critical to account for the observed
correlation decay. Alternative \#1
for our purposes may thus be termed SOC. On the other hand, in alternative \#2
the potentials are
non-summable (= long-ranged), so that this case corresponds to the SOLRI
scenario. To support the latter, we make
some key comparisons with long-ranged systems.

An analogy was already remarked a few years ago between the DDS and dipolar
systems \cite{ZS}. Of course, dipole
spin-spin correlations are also $\propto r^{-d}$ even at high-temperature. This
is consistent with our point of view,
because the dipole potential just misses being absolutely summable (and is thus
``non-Gibbsian'' according to
our criterion!) An important consequence of this non-summability was early
recognized \cite{VOS}, namely, that the
thermodynamics of dipolar systems is {\em shape-dependent}. This situation
arises because the dipole potential energy
sums are only conditionally convergent and hence may lead to different values
depending upon the order of summation,
i.e. the shape, at least at nonzero field \cite{Griff}. In uniformly-magnetized
(=high field), ellipsoidally-shaped
samples of dipolar materials the shape dependence of thermodynamic free-energy
functions is simply parameterized \cite{LH},
in good agreement with experiment \cite{LL}.

It was argued in \cite{ELS} that a similiar shape-dependence occurs in DDS. The
thermodynamic functions of interest
are the ``pressure'' $p_\bE$ and the ``Helmholtz free-energy'' $f_\bE$. The
former is defined by the thermodynamic limit
\be p_\bE(\mu,\beta|\rho_*)= \lim_{\Lambda\rightarrow\bZ^d}{{1}\over{\beta
|\Lambda|}}\log\langle
e^{\beta\mu N_\Lambda}\rangle_{\rho_*,\beta,\bE}, \lb{Gibbs} \ee
where $\Lambda$ is a sequence of lattice volumes converging to $\bZ^d$,
$N_\Lambda(\eta)$ is the number of
particles within $\Lambda$ for the configuration $\eta$, and the average
$\langle\cdot\rangle_{\rho_*,\beta,\bE}$
is with respect to the invariant measure $\mu_{\rho_*,\beta,\bE}$ of the DLG
for reference density $\rho_*$.
The ``Helmholtz free energy'' $f_\bE$ is then introduced by the Legendre
transform $f_\bE(\rho,\beta|\rho_*)
=\sup_\mu[\mu\rho-p_\bE(\mu,\beta|\rho_*)]$. We include the $\rho_*$ as a
reminder of the reference
density and the $\bE$ to indicate the strength of the applied electric field.
Since the measures here are for
irreversible steady-states with Ohmic dissipation, these are not usual free
energies. They coincide with the
equilibrium free-energies in the limit $E\rightarrow 0$. The physical
interpretation of $f_\bE(\rho,\beta|\rho_*)$
is as an ``excess dissipation function,'' i.e. as the total energy dissipated
per volume by an external field to
change the density to $\rho$ from its reference value $\rho_*$ (in addition to
the Ohmic dissipation intrinsic to
the reference state) \cite{ELS}. The argument for shape-dependence is that the
``susceptibility'' (essentially, the
isothermal compressibility) may be written both in terms of the free-energy,
$\chi_\bE(\rho,\beta) = \left[\beta
{{\partial^2 f_\bE}\over{\partial \rho^2}}(\rho,\beta)\right]^{-1}$, and also
in terms of the structure-function,
via the limit $\chi_\bE=\lim_{k\rightarrow 0}\wS_\bE(\bk)$. However, the latter
limit is indeterminate when the structure
function has the form in Eq.(\ref{stat-struc}) and depends upon the wavenumber
vector direction $\widehat{\bk}$
along which the limit is taken. Thus, the free-energy itself must be
shape-dependent, by the same argument as
for dipole systems.

To test this prediction we have performed a Monte Carlo simulation of the DLG
on a periodic square $S\times S$ lattice with
Ising Hamiltonian $H= - \frac{1}{2}\sum_{\langle \bx,\by
\rangle}\eta_\bx\eta_\by$ where $\langle\bx,\by\rangle$ denotes
nearest neighbor sites, for which the (inverse) critical temperature is
$\beta_c\approx 0.31\cite {SZ}$. To stay well
within the single phase (high temperature) regime, we used $\beta=0.2$,
$E=10.0$ and reference density $\rho_*=0.5$.
We have determined the thermodynamic functions for rectangular subblocks
$\Lambda$ of the $S\times S$ system, in which
various aspect ratios of the sides of the rectangles were chosen. The pressure
was evaluated by a double limit. First the
infinite volume limit was obtained by a linear extrapolation in $1/S\rightarrow
0$ on the Monte Carlo average $\langle
e^{\beta\mu N_\Lambda}\rangle_S$ in the steady-state with $S=64,128,256,512$.
The thermodynamic limit in Eq.(\ref{Gibbs})
for the pressure  was then evaluated by a second linear extrapolation in the
inverse volume $1/|\Lambda|$ of the subblock
going to zero. The largest subblock edge in this second extrapolation had
length 18. The Legendre transform to the
free-energy was then carried out. The results are shown in Figure 1 for
subblocks with aspect ratios of 3:1 and 1:3 for
sidelengths parallel and perpendicular to the field, respectively. Error bars
reflect both statistical deviations in
independent runs and the double extrapolation procedure. The two functions are
clearly distinct. We see that the DLG,
considered as a model of a current-carrying electrochemical cell, has
well-defined free-energies but the results depend
upon the shape of the cell!

This shape-dependence leads, however, to a rather serious puzzle about the
hydrodynamic behavior of the DDS.
The general problem is to describe how an initial smooth density profile
relaxes to a constant density
in the driven steady-state. In \cite{ELS} a nonlinear hydrodynamic equation was
derived by the formal method of
nonequilibrium distributions to describe this irreversible process. Density
fields varying on a length-scale of
the order of $\en^{-1}$ compared with the lattice distance were formally shown
to evolve by a {\em drift-diffusion equation},
\be \dot{\rho}(\br,t) =
-\grad\bdot\left[\en^{-1}\wbj_\bE(\rho)-\beta\bL_\bE(\rho)\bdot\grad
\left({{\delta{\cal F}_\bE}\over{\delta\rho}}\right)\right], \lb{GL-eq} \ee
over times of order $\en^{-2}$.
This equation has the ``Onsager form'', with $\bj_\bE(\rho)$ the conduction
current, $\bL_\bE(\rho)$ the Onsager
coefficient matrix, and ${\cal F}_\bE[\rho]$ the free-energy functional.
Explicit analytical formulae were given
in \cite{ELS} for each of these quantities, e.g. a Green-Kubo formula for the
Onsager matrix. These formulae are
exact even at high field strengths $E$, although they may be difficult to
evaluate concretely.
The free-energy functional in \cite{ELS} was {\em nominally} given by the local
expression
${\cal F}_\bE[\rho]=\int d\br f_\bE(\rho(\br))$. However, as observed there,
such a form is indeterminate.
Since the free-energy depends upon the limiting shape, which value is to be
used?

We can now resolve this issue. The free-energy functional actually shown in
\cite{ELS} to be relevant to hydrodynamics
is given by a Legendre transform $ {\cal F}_{\bE}[\rho]=\int d\br
\rho(\br)\mu(\br)-{\cal P}_{\bE}[\mu]$
of the pressure functional
\be {\cal P}_{\bE}[\mu]= {{1}\over{\beta}}\lim_{\en\rightarrow 0}
                        \en^d\log\langle\exp[\sum_\bx \beta\mu(\en
\bx)\eta_\bx]\rangle_{\rho_*,\beta,\bE}.
\lb{press-func} \ee
Simple computation then gives $\left.{{\delta {\cal
P}_\bE}\over{\delta\mu(\br)}}\right|_{\mu=0}=\rho_*$ and
\begin{eqnarray}
\left.{{\delta^2 {\cal
P}_\bE}\over{\delta\mu(\br)\delta\mu(\br')}}\right|_{\mu=0} & = & \beta
     \lim_{\en\rightarrow 0}
\en^{-d}[\langle\eta_{[\en^{-1}\br]}\eta_{[\en^{-1}\br']}
\rangle_{\rho_*,\bE}-\rho_*^2] \cr
 \, & \equiv & \beta S_\bE(\br-\br'). \lb{2nd-deriv}
\end{eqnarray}
It follows that ${\cal P}_{\bE}[\mu]= \int d\br \rho_* \mu(\br) +
{{\beta}\over{2}}\int d\br\int d\br'
S_\bE(\br-\br')\mu(\br)\mu(\br') + O(\mu^3)$, and the Legendre transform yields
\be {\cal F}_{\bE}[\rho]= {{1}\over{2\beta}}\int d\br\int d\br'
S^{-1}_\bE(\br-\br')\delta\rho(\br)\delta\rho(\br')
+ O(\delta\rho^3) \lb{Helm-nloc} \ee
where $\delta\rho(\br)\equiv \rho(\br)-\rho_*$ and $S^{-1}_\bE$ is the operator
inverse of $S_\bE$. In other words,
$\widehat{S^{-1}_\bE}(\bk)=\bk\bdot\bD_E\bdot\bk/\bk\bdot\bL_E\bdot\bk$. Hence,
the inverse kernel $S^{-1}_\bE(\br)$
is $\propto r^{-d}$ for large $r$, too.  We see that the hydrodynamic equation
of the DDS at finite field strengths $E$ thus must have an explicit, severely
{\em non-local} form. [H. Spohn has
emphasized to us that Eq.(\ref{GL-eq}) linearized about the homogeneous state
of density $\rho_*$ will
still be local if Eq.(\ref{stat-struc}) holds, since then
$\bL_\bE(\rho_*)\bddot\grad\grad\int d\br'
S^{-1}_\bE(\br-\br')\delta\rho(\br')= \bD_\bE(\rho_*)\bddot\grad\grad
\delta\rho(\br)$.] However, with the free-energy
functional in (\ref{Helm-nloc}) replacing the local expression, all results of
\cite{ELS} remain valid: the H-theorem,
the fluctuation-dissipation theorem, etc.

Such nonlocal hydrodynamic behavior should also be present in equilibrium
systems with long-ranged interactions,
e.g. dipolar hard sphere systems or ferrofluids.  In one such system, the
Kawasaki lattice gas with a long-ranged
Kac pair-potential, there is a rigorous result \cite{GL} that a hydrodynamic
equation of the same form as
(\ref{GL-eq}) holds, with a similar nonlocal expression for the free-energy as
in (\ref{Helm-nloc}),
simply replacing $S_\bE^{-1}(\br)$ by the kernel $J(\br)$ of the Kac
potential.We see again a very striking
and fruitful analogy between dissipative, driven systems and equilibrium
systems with long-range interactions.

Our results verify that the SOLRI scenario holds in our model, and not SOC. For
Gibbsian measures with summable potentials
there is no shape-dependence of thermodynamics, such as we observe here, even
at the critical point (\cite{Israel}, Theorem I.2.5). Boundary
conditions play a role {\em below} the critical point in the phase coexistence
region in determining which of multiple
phases will occur, but even then the free energies are independent of the
phase. The ordinary thermodynamic limit, with
no shape-dependence, remains valid directly at the critical point. Likewise,
the dynamics of short-ranged systems in the
coexistence region is expected to be described by the Cahn-Hilliard dynamics,
which is local. It is actually a little perplexing
how to interpret the SOC point of view that nonequilibrium steady-states are
always ``critical'', when these are observed
themselves to undergo continuous phase-transitions at sharp values of
temperature and/or density. Such transitions occur
both in the DLG \cite{SZ} and in granular flow \cite{Will}, not to mention
dipole systems \cite{AF}. Away from the transition point
both DDS and dipole systems have a finite correlation length $\xi$ which
characterizes the crossover from a critical
power-law $\propto r^{-(d-2+\eta)}$ at intermediate range $r\ll \xi$ into the
asymptotic power-law $\propto r^{-d}$ at
long-range $r\gg \xi$ \cite{SZ,AF}.

It is most interesting to consider the implications of our model calculation
for real systems. We expect that the main
conclusions concerning shape-dependence and nonlocality will hold for physical
drift-diffusion systems, such as fluid
or solid electrolytes, semiconductors, and, at a more mesoscopic level,
colloidal suspensions \cite{Feld}. In real
charged-particle systems there is the added complication of a dynamics which is
itself long-ranged, via Coulombic
interactions. However, these are expected to be Debye-screened and effectively
short-ranged. The drift-diffusion equations
were justified long ago for nonequilibrium processes in electrolyte solutions
at low density and small fields within
Debye-H\"{u}ckel theory \cite{OF}. The equations are of the same form as those
we have considered, simply generalized
to multiple ionic species. The effects of the Coulomb interactions are
calculable and can all be incorporated into an
effective Onsager matrix, with cross-species terms due to ionic-cloud
distortion and electrophoresis. The effects
considered in our work correspond to contributions to the invariant measures
and thermodynamic potentials in at least
the $E^2$ power of the field strength. It would be interesting to make a
theoretical estimate of the order of magnitude
of such effects in electrolyte solutions. Perhaps the most accessible
predictions are the long-ranged correlations
themselves, which could be observed in light-scattering experiments similiar to
those carried out on
simple fluids subject to a temperature gradient (see \cite{DKS} for a recent
review). Also of possible practical
interest are the implications of a nonlocal hydrodynamics for granular flow.

\noindent {\bf Acknowledgements:} The authors wish to thank B. M. Boghosian, H.
Gould, W. Klein, J. L. Lebowitz,
Y. Oono and H. Spohn for helpful discussions. FJA was funded in part by NSF
Grant DMR 9633385 and by AFOSR Grant
\#F49620-95-1-0285.

\begin{center}
FIGURE CAPTION
\end{center}

Figure (1.)  Free energy as a function of density for $E=10$,
$\beta=0.2$, $\rho_*=0.5$ for aspect ratio 3:1 (triangles) and
1:3 (squares).

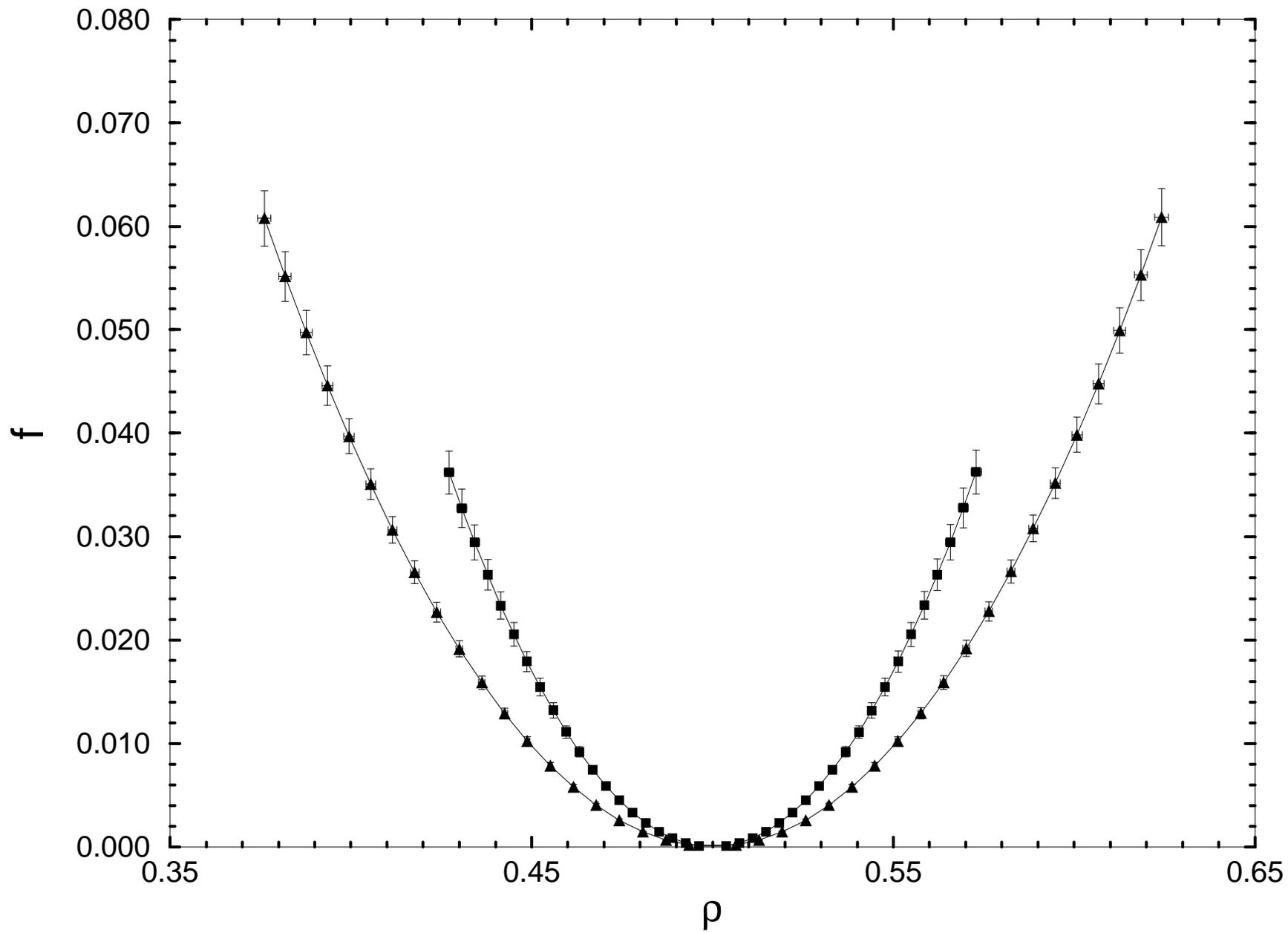
\end{document}